\documentstyle[prl,aps,preprint,tighten,floats,aps,epsfig,psfig]{revtex}

\newcommand{\beq}{\begin{eqnarray}}
\newcommand{\eeq}{\end{eqnarray}}
\newcommand{\eq}{\ref}

\newcommand{\MSB}{\overline{\rm MS}}

\newcommand{\Lams}{\Lambda_{\overline{\rm MS}}}

\newcommand{\be}{\begin{equation}}
\newcommand{\ee}{\end{equation}}
\newcommand{\lwrsim}{\raise0.3ex\hbox{$<$\kern-0.75em\raise-1.1ex\hbox{$\sim$}}}
\newcommand{\bk}{(\beta,\kappa_{\rm sea})}

\def\eq#1{eq. (\ref{#1})}
\def\jhep#1#2#3{J. High Energy Phys. {\bf #1} (#2) #3}
\def\prd#1#2#3{Phys.\ Rev.\ {\bf D#1} (#2) #3}
\def\npb#1#2#3{Nucl.\ Phys.\ {\bf B#1} (#2) #3}
\def\nppb#1#2#3{Nucl.\ Phys.\ Proc. \ Suppl. \ {\bf B#1} (#2) #3}
\def\nppa#1#2#3{Nucl.\ Phys.\ Proc. \ Suppl. \ {\bf A#1} (#2) #3}
\def\plb#1#2#3{Phys.\ Lett.\ {\bf B#1} (#2) #3}

\def\prl#1#2#3{Phys.\ Rev.\ Lett.\ #1 (19#3) #2}

\def\epjc#1#2#3{Eur. \ Phys. \ J. \ {\bf C#1} (#2) #3}

\begin{document}

\title{ Preliminary Calculation of $\alpha_s$ 
from Green Functions with Dynamical Quarks.}
\author{ Ph. Boucaud$^a$,  J.P. Leroy$^b$, J. Micheli$^b$, H. Moutarde$^c$ \\  
O. P\`ene$^b$, J. Rodr\'\i guez--Quintero$^d$,  
 and C. Roiesnel$^c$ } \par \maketitle
\begin{center}
$^a${\sl Dip. di Fisica, Univ. di Roma ``La Sapienza'' and INFN, Sezione di Roma,
\\ Piazzale Aldo Moro 2, I-00185 Rome, Italy}
\\$^b${\sl Laboratoire de Physique Th\'eorique~\footnote{Unit\'e Mixte 
de Recherche du CNRS - UMR 8627}\\
Universit\'e de Paris XI, B\^atiment 211, 91405 Orsay Cedex,
France}
\\$^c$ {\sl Centre de Physique Th\'eorique\footnote{
Unit\'e Mixte de Recherche C7644 du CNRS} de l'Ecole Polytechnique\\
91128 Palaiseau cedex, France}
\\$^d${\sl Dpto. de F\'{\i}sica Aplicada \\
E.P.S. La R\'abida, Universidad de Huelva, 21819 Palos de la fra., Spain}  

\end{center}

\begin{abstract}
 We present preliminary results on the computation of the QCD running coupling
constant in the $\widetilde{MOM}$ scheme and Landau gauge with two flavours of
dynamical Wilson quarks. Gluon momenta range up to about 7 GeV ($\beta =$ 5.6,
5.8 and 6.0) with a constant dynamical-quark mass. 
This range already allows to
exhibit some evidence for a sizable $1/\mu^2$ correction to the asymptotic
behaviour, as in the quenched approximation, although a fit without power
corrections is still possible with a reasonable $\chi^2$.
Following the conclusions of our quenched study, we take into account $1/\mu^2$ correction to the asymptotic behaviour. We find $\Lambda_{\rm
\overline{MS}}^{N_f=2} = 264(27) {\rm MeV } \times [{a^{-1}(5.6,0.1560)}/{2.19\,
{\rm GeV}}] $, which leads to $\alpha_s(M_Z) = 0.113(3)(4)$. The latter result has
to be taken as a preliminary indication rather than a real prediction in view of
the systematic errors still to be controlled.  Still, being two sigmas below the
experimental result makes it very encouraging.

\end{abstract}
\begin{flushright} LPT Orsay-01-70\\
FT-UHU/01-04\\  
CPhT-S033.0701 \\
\end{flushright}
\newpage

The non-perturbative calculation of the running coupling constant of 
QCD is certainly a very important problem. 
In pure Yang-Mills it has been performed 
with several different methods, the most systematic ones 
using the Schr\"odinger functional \cite{luscher}, 
and the gluon Green functions \cite{alles,nous,direnzo,nouspdeux}.  
It is noticeable that the latter two methods, although very
different, end up with perfectly compatible values for $\Lambda_{\rm QCD}$.

Of course the real challenge is to compute $\alpha_s$ with dynamical
fermions.  This task has been undertaken using NRQCD  several years   ago
\cite{khadra,davies,SESAM} and, once extrapolated to $M_Z$ 
 leads to rather satisfactory values for $\alpha_s(M_Z)$. 
 Recently, the  QCDSF-UKQCD collaboration \cite{schierholz} and
 the ALPHA one \cite{sommer} have reported progress in determining
 $\alpha_s$ with two flavours using relativistic lattice QCD
 and nonperturbatively improved Wilson fermions.
  
 In this letter we will report our work consisting in applying  the Green
function method estimate \cite{alles,nous,direnzo,nouspdeux} with non-improved
Wilson dynamical quarks.   The principle of the method is quite simple since
it consists  in following the steps which are standard in perturbative QCD in
the momentum subtraction scheme. This gives immediately a nonperturbative
estimate of the coupling constant at different scales. Its running can be
confronted to the perturbative QCD expectation. We use the $\widetilde {MOM}$ 
renormalization  scheme which corresponds to using an
asymmetric subtraction point :  $p_1^2=p_3^2\equiv\mu^2, p_2=0$. This scheme
proved to give rather good signals and, in spite of the zero momentum,  
no pathology has been seen. 

From our study of the pure Yang-Mills case \cite{nous,direnzo,nouspdeux}
 we have learnt two
main lessons: one is that a study of the asymptotic  behaviour of 
$\alpha_s$ needs a large energy window, since  the value
of $\Lambda_{\rm QCD}$ we are looking for  depends on the weak logarithmic dependence of
$\alpha_s$ on the energy scale $\mu$, the second is that the $1/\mu^2$ correction
can be sizable up to a large  energy.

We aim at computing $\alpha_s$, $\Lambda_{\rm QCD}$ and the power correction
term with two flavours of dynamical quarks. This requires, as we shall see in
more details, an exploration of the two-dimensional ($g_0, m_{\rm sea}$) bare
parameter space. To this goal we have run lattice simulations on several
$16^4$ lattices. Notwithstanding the modest volumes of these lattices, we
realised that some interesting  physics can already be extracted. Furthermore, the exploration of the bare parameter space provides us with new data which might be of interest for unquenched studies by the lattice community.
 This legitimates in our opinion a progress report which is the
 aim of this paper.
  
 \section{Our strategy}
  \label{strategy}

We have computed in the Landau gauge the two-gluon and
three-gluon  Green functions  leading to a nonperturbative calculation of  
$\alpha_s^{\rm Latt}(\mu)$ in  the well defined MOM schemes \cite{alles,nous}.
At energies $\mu$  above 2.6  GeV we will fit this function by
\beq
\alpha_s^{\rm Latt}(\mu^2)=\alpha_{\rm s,pert}
(\mu^2)\,\left(1+ \frac{c}{\mu^2}\right) \ ,
\label{LaFor}
\eeq
where  $\alpha_{\rm s,pert}(\mu^2)$
is the perturbative running coupling constant computed to four loops
from some fitted $\Lambda_{\rm QCD}$, and  $\alpha_{\rm s,pert}\, c/\mu^2$ is a
 power correction
which has proven, in the $N_f$~=~0 case, not to be negligible  up to 10 GeV, and was
eventually traced back to an OPE condensate \mbox{$<A_\mu A^\mu>$.}
The reason for
 choosing as in eq.  (\ref{LaFor}) a non perturbative
 correction $\propto  \alpha_{\rm s,pert}(\mu^2) /\mu^2$ instead
 of simply  $\propto 1/\mu^2$ is twofolds: 
 \begin{description}
 \item[ (i)] Theoretically, an OPE study~\cite{nouspdeux} including a computation of the anomalous 
 dimension of the coefficient of $<A^2>$  leads to 
 an expected energy dependence close to $\alpha_{\rm s,pert}(\mu^2)/\mu^2$;
\item[(ii)] Practically in the quenched as well as unquenched case
 the fit with $\alpha_{\rm s,pert}(\mu^2)/\mu^2$
 is much more stable for changes of the energy window
than the fit with $1/\mu^2$. The former stability will be illustrated   
in table \ref{table_lambda}. 
\end{description}

Interestingly, $\alpha_s^{\rm Latt}$
is at the same time both the goal of our study and a very useful
tool: from the lattice simulations one extracts the continuum $\alpha_s$
up to small lattice artifacts; lattice spacing ratios are then fitted 
to preserve the continuity of $\alpha_s(\mu)$ for the whole set of data.   

This  program is performed on hypercubic lattices in order to simplify the necessary
 tensorial analysis
of the Green functions\footnote{ This does not allow to compute the $\rho$ 
meson mass, 
which is better performed on lattices longer in time direction than in space.
Some consequences will be discussed  later.}.  In the $N_f=0$ case
we  combined $\beta=6.0, 6.2, 6.4,
6.8$ quenched lattice simulations,  i.e. a lattice spacing ranging from
$\sim 0.03$ fm to $0.1$ fm, in \cite{nous,nouspdeux}, allowing
to reach momenta up to 10 GeV.

  With dynamical
fermions the physics depends on two parameters, $\beta$  which represents the
bare coupling constant  and $\kappa_{\rm sea}$ representing the bare dynamical-quark
mass.  A wide energy window is reached by combining simulations with different
lattice spacings and the same {\it renormalised dynamical-quark mass expressed in
physical units}. The problem is of course that we  do not know a priori for a
given $\beta$ which $\kappa_{\rm sea}$  corresponds to one given renormalised
dynamical-quark mass in physical units. This needs as mentioned above some
exploration of the $\bk$  parameter space to find one or several lines of equal
dynamical masses. In view of the computational
cost of such an exploration we have chosen to perform it on a small volume,
$16^4$. 

We now would like to sketch our strategy to compute 
the lattice spacings and the renormalised dynamical-quark masses in this
exploratory stage on a $16^4$ volume.
 We  proceed as follows.  We start from a calibrating set of parameters ($\beta, \kappa_{\rm
sea}$) for which some published results yield the inverse lattice spacing $a^{-1}$ computed
from some hadronic quantity, for example the $\rho$ meson mass. We then estimate $a^{-1}$ for
other values of ($\beta, \kappa_{\rm sea}$) by matching\footnote{Our matching procedure was proved to be succesful when applied to quenched data, where lattice spacings are well known \cite{bali}.} the value of $\alpha_s(\mu)$. This uses
as an assumption that  we may neglect the dependence of $\alpha_s$ on the dynamical-quark
mass, at least in the mass range under consideration. This assumption is not more 
arbitrary than any other
calibration based on, for example, the physical $\rho$ meson mass, which neglects the unknown
dependence of the $\rho$ meson mass on the dynamical-quark mass.

Once we have estimated the lattice spacings for all our lattices
with different sets ($\beta,\kappa_{\rm sea}$), we estimate $m_{\rm sea}$ 
from the ratio $\partial_\mu A^\mu/P_5$ where 
$A^\mu$ is the axial current and $P_5$ the pseudoscalar density, computed for
a valence quark\footnote{We call valence quarks the quarks which contribute
to the current densities and propagate in the gauge field background. The latter
depends on the sea quark mass. The theory is unitary only
if valence and sea quarks have the same mass. However we will also make use of
 $\kappa_{\rm val} \ne \kappa_{\rm sea}$ as an intermediate step.} with the same bare 
mass as the dynamical quark ($\kappa_{\rm val} = \kappa_{\rm sea}$).

At the end of this procedure we can fix with a reasonable accuracy
a set of couples ($\beta,\kappa_{\rm sea}$) which includes our calibrating
 lattice and for which $\beta$ the  dynamical-quark mass remains constant in physical
 units when  $\beta$ is varied.
 This knowledge allows for a preliminary analysis 
of $\alpha_s$ with two flavours, of the resulting 
$\Lambda_{\overline {\rm MS}}^{N_f=2}$
and power correction term,  and finally of $\alpha_s(M_Z)$.
This will be presented in this letter.

Still we do not forget that finite volume effects may be large in such a small
volume, that our dynamical-quark masses are large, etc. We will therefore briefly discuss sources of systematic uncertainties at the end of this letter. We are however now in a
position to launch the calculations on a larger volume $24^4$ and/or with lighter
masses and correct for the biases of our present results.  

\section{Some useful perturbative formulae}
We now proceed to establish the conventions and to introduce the formulae 
that we will use in the following. $\alpha_{\rm s,pert}(\mu^2)$ in \eq{LaFor} 
stands for the perturbative running coupling constant expanded up to 
the fourth loop and verifying (in this section we write $\alpha$ instead of 
$\alpha_{\rm s,pert}(\mu^2)$ to simplify the notations)
\beq
\frac{d}{d\ln{\mu}}\alpha \ = \ - \left( \frac{\beta_0}{2 \pi}
\alpha^2 + \frac{\beta_1}{4 \pi^2} \alpha^3 + \frac{\beta_2}{64 \pi^3} 
\alpha^4 + \frac{\beta_3}{128 \pi^4} \alpha^5 \right).
\label{beta}
\eeq

In all schemes
\beq
\beta_0 = 11 -  \frac 2 3 {N_f}\quad \beta_1 = 51 - \frac {19}3 {N_f} \ ,
\eeq
while $\beta_2, \beta_3$ depend on the particular scheme. The values we need for the $\widetilde{\rm{MOM}}$  scheme can be found in ref.~\cite{chet}.
\noindent  The exact integration of \eq{beta} to the third loop, with 
the standard boundary condition defining the $\Lambda$ parameter~\cite{pdg}, 
leads to~\cite{nous}

\[
 \Lambda^{3 {\rm loops}} =\Lambda^{(c)}(\alpha)\left(1+\frac 
 {\beta_1\alpha}{2\pi\beta_0}+
 \frac
 {\beta_2\alpha^2}{32\pi^2\beta_0}\right)^{\frac{\beta_1}{2\beta_0^2}}
  \]
  \beq \times \exp
\left\{\frac{\beta_0\beta_2-4\beta_1^2}{2\beta_0^2\sqrt{\Delta}}\left[
\arctan\left(\frac{\sqrt{\Delta}}{2\beta_1+\beta_2\alpha/4\pi}\right)
-\arctan\left(\frac{\sqrt{\Delta}}{2\beta_1}\right)\right]\right\}
 \label{lambda3}\eeq

\noindent where $\Lambda^{(c)}$ denotes the conventional two loops formula:
 \beq       
             \Lambda^{(c)} \equiv \mu \exp\left (\frac{-2 \pi}{\beta_0
	      \alpha}\right)\times
	      \left(\frac{\beta_0  \alpha}{4 \pi}\right)^{-\frac {\beta_1}
	      { \beta_0^2}}\label{lambda} \;\; ;
\label{LambdaConv}
\eeq

\noindent and $\Delta\equiv 2\beta_0\beta_2-4\beta_1^2>0$ 
in the $\widetilde{\rm{MOM}}$  scheme which we use. If one only retains the first correction 
coming from the perturbative fourth loop, it can then be written

\beq
\Lambda^{4 {\rm loops}} \ = \ \Lambda^{3 {\rm loops}} \ 
\exp\left(-\frac{\beta_3}{64 \pi^2 \beta_0^2} \alpha^2 \right) \ \ .
\label{lambda4}
\eeq.

 In the previous formula, of course, the use of  $\Lambda$, $\alpha$ and
$\beta$'s  stands for the $\Lambda$ parameter, the running  coupling constant and
beta function coefficients in the  particular  $\widetilde{\rm{MOM}}$
renormalisation scheme. From now on we will  systematically convert $\Lambda$
into $\Lams$ using 
\cite{nous}
\beq
\Lams = \Lambda \,  \exp \left[-\frac{1}{22} \left(  \frac{70}{3} - \frac{22}{9} N_f \right) \right]
\; \label{raplambdas}
\eeq

No analytical expression can exactly inverse neither three-loop 
\eq{lambda3} nor four-loop \eq{lambda4}. The following formula gives nevertheless
an approximated solution to the inversion of the perturbative expansion of 
\eq{lambda4}: 

\begin{eqnarray}
\protect\label{3lp}
\alpha_{\rm s,pert}(\mu^2) & = & \frac{4 \pi}{\beta_0 \, t} \, - \, 
\frac{8 \pi \beta_1}{\beta_0} 
\frac{\log(t)}{(\beta_0 \, t)^2} \nonumber \\
&+ &  \, \frac{1}{(\beta_0 \, t)^3} \, 
\left( \frac{2 \pi \beta_2}{\beta_0} + 
\frac{16 \pi \beta_1^2}{\beta_0^2} 
(\log^2(t) - \log(t) - 1 ) \right) + \frac{1}{(\beta_0 t)^4} \nonumber \\
&\times &  \,
\left[\frac{2 \pi \beta_3}{\beta_0} + 
  \frac{16 \pi \beta_1^3}{\beta_0^3} 
\left( -2 \log^3(t) + 5 \log^2(t) + \left( 4 - \frac{3\beta_2 \beta_0}{4\beta_1^2} \right) 
\log(t) - 1 \right)
\right] \nonumber \\
\label{alpha4}
\end{eqnarray}

\noindent where $t = \log(\mu^2/\Lambda^2)$.  The exact numerical inversions
of, for instance, \eq{lambda4}, can be easily obtained;  but, of course,
such an exact inversion and the approximated solution in \eq{alpha4}  should
only differ by perturbative contributions of order higher than four loops.

\section{The First iteration}
\label{calculation}

\subsection{Lattice spacings}

The lattice parameters which we have used for our simulations 
are displayed in 
table \ref{table_calc} together with our estimates of the pseudocritical 
$\kappa$'s, $\kappa_{pc}$, defined in subsection \ref{pc}, of the  
 the lattice spacings and of the sea quark masses. 

	We will not repeat the method used to extract 
$\alpha_s$ from Green functions as it is exactly similar to what was done in
the pure Yang-Mills case. The $\widetilde{\rm{MOM}}$
 scheme uses the  ``asymmetric'' three point
Green function, i.e. with gluon squared momenta ($0, \mu^2, \mu^2$) where
$\mu^2= n(2 \pi/L)^2$,  $n$ being an integer\
\footnote{Any integer verifies $n=\sum_{i=0,3} n_i^2$ for at least one set
of integers $n_i,i=0,3$ (Lagrange's four-square theorem).}. Let us nevertheless recall that the ``asymmetric'' three-point Green function turned out to be more convenient for our purpose than the symmetric one (i.e. with gluon square momenta ($\mu^2, \mu^2, \mu^2$)): the high accuracy achieved in our quenched study \cite{alles,nous,direnzo,nouspdeux} gave us some evidence that the ``asymmetric signal'' was less noisy than the symmetric one, while being as reliable (more momenta available and no observable infrared pathology due to the zero momentum).

\vskip 0.5 cm
\begin{table}
\begin{center}
\begin{tabular}{|c|c|c|c|c|c|}\hline
$\beta$ & $\kappa_{\rm sea}$ & Volume & $\kappa_{pc}$ &$a^{-1}$ (GeV) &
 $m_{\rm sea}$ (MeV)
\\ \hline
5.6 \cite{SESAMold} & 0.1560 & 16$^3 \times 32$ &  & 2.19(8) & \\ 
5.6 \cite{SESAMold} & 0.1575 & 16$^3 \times 32$ &  & 2.38(7) & \\  
5.6 \cite{FAC} & 0.1575 & 24$^3 \times 40$ & 0.15927(5) & 2.51(6) & \\ 
5.6 \cite{FAC} & 0.1580 & 24$^3 \times 40$ &  0.15887(4) & 2.54(6) & \\
\hline
\hline 
 5.6 & 0.1560 & 24$^4$ & 0.16053(3) & 2.19(8) &  164(7)\\
 5.6 & 0.1560 & 16$^4$ & 0.16048(13) & 2.19(8) & 164(7)\\
 5.6 & 0.1575 & 16$^4$ & 0.1593(1) & 2.42(9) & 79(3) \\
 \hline
 5.8 & 0.1500 & 16$^4$ & 0.15672(6) & 2.45(13) & 325(18)\\
 5.8 & 0.1525 & 16$^4$ & 0.15555(12) & 2.76(7) & 173(4)\\
 5.8 & 0.1535 & 16$^4$ & 0.15522(9) & 2.91(18) & 103(16)\\
 5.8 & 0.1540 & 16$^4$ & 0.15499(6) & 3.13(13) & 64(5)\\
 \hline
 6.0 & 0.1480 & 16$^4$ & 0.15272(7) & 3.62(10) & 391(12)\\
 6.0 & 0.1490 & 16$^4$ & 0.15262(7) & 3.73(13) & 308(12)\\
 6.0 & 0.1500 & 16$^4$ & 0.15238(4) & 3.78(14) & 213(3)\\
 6.0 & 0.1505 & 16$^4$ & 0.15240(5) & 3.84(15) & 169(8)\\
 6.0 & 0.1510 & 16$^4$ & 0.15207(3) & 3.96(16) & 96(4)\\
 \hline
\end{tabular}
\vskip 1.5 cm
\caption{ Data taken from literature and first iteration estimates from our runs.
$\kappa_{pc}$ is defined in subsection \ref{pc}. The dynamical-quark masses are renormalised in the $\MSB$ scheme at 3 GeV.} 
\label{table_calc}
\end{center}
\end{table}

In table \ref{table_calc}, we give the full set of runs performed and the
preliminary values obtained for $a^{-1}$ and  $m_{\rm sea}$. In the   first four
rows, we also quote sets of values taken from  literature
\cite{SESAMold,FAC}. In particular, the value $a^{-1} = 2.19(8)$ GeV for 
(5.6, 0.1560) is taken from \cite{SESAMold}  and we will use it to calibrate all
our runs. As already mentioned, ratios  of lattice spacings result from imposing
the continuity  of $\alpha_s(\mu)$ from different lattices, neglecting the
expected small dependence of $\alpha_s$ on the dynamical mass  $m_{\rm sea}$.
Since the error on the calibrating $a^{-1}$ propagates trivially to 
$\Lambda_{\rm QCD}$ we will use 2.19 GeV without its error until eq.
(\ref{lambdanf0}); a discussion of these errors will follow eq. (\ref{lambdanf0}).
 Thus the errors quoted here for $a^{-1}$ only stand for the
ratios.

\subsection{Sea-quark masses}

Once the lattice spacings are estimated, we also need to compute
$a m_{\rm sea}$. To this aim we compute the propagators 
of valence quarks for several $\kappa_{\rm val}$ among which
one with $\kappa_{\rm val} = \kappa_{\rm sea}$ in order
to be able to deduce the mass of the sea quark from the
 estimated mass of the valence quark. 
  
This is done using the ratio 
\beq
\rho = \frac 1 2\frac {\sum_{\vec x} P_5(0)\partial_0 A_0(\vec x, t) }
{\sum_{\vec x} P_5(0)P_5(\vec x, t)}
\label{rho}
\eeq
where $P_5$ is the pseudoscalar density, and $A_\mu$ the axial
current. 

To estimate the ratio $\rho$ in  (\ref{rho}) we have used two methods. 
The simplest consists in looking for a plateau of the ratio, the time derivative
in the numerator being computed by a symmetrised discrete difference.  However this 
``brute-force'' estimate appeared to be affected by some strong $O(a)$ effects in several cases.

The second method, which is a variant of the one proposed in \cite{CPPACS}, fits
on some time interval the $<P_5P_5>$ in the denominator by a $\cosh$ function and
the $<P_5A_0>$ in the numerator by a $\sinh$ with the same ``mass'' 
term\footnote{It
 is not really a mass since 
 the very short time interval considered does not allow to isolate
 the ground state. It nevertheless turns out that the data for 
 $5 \le t  \le 11$ can be satisfactorily fitted respectively with a $\cosh$
 and a $\sinh$.}.  The time
derivative of the $\sinh$ in the numerator is then proportional to the $\cosh$ in
the denominator and  the ratio gives an estimate of the ratio $\rho$ in
(\ref{rho}). We  have used the second method because it  turned out to be more 
stable against the change of parameters (domain of the fit)
and to provide a better continuity when $\kappa$ is changed.

The valence mass is given by 
\beq
a m_{\rm val} = \frac {Z_A} {Z_P} \rho
\eeq
For $Z_A$ and $Z_P$ in the RI-MOM scheme, we have taken \cite{damir}  $Z_A=0.77(1)$ and
$Z_P=0.54(1)$ i.e.  $Z_A/(2Z_P)\simeq 0.71$ at $\mu=3$ GeV (the value of $Z_P$ is derived from the Ward
identity value of  $Z_P/Z_S$~\cite{vladikas}). The large Goldstone pole contribution  stressed in
ref. \cite{Cudell} is claimed\footnote{Indeed, in ref. \cite{damir2} this 
method is applied and agrees with ref. \cite{sommer}, obviously free of Goldstone
boson contribution. Although the Goldstone boson question is not commented in \cite{damir2}
this seems to confirm the above-mentioned claim.} to be eliminated in this value of  $Z_P$. A more careful study of
the renormalization constant will be performed soon. 

From $a m_{\rm sea}$ and $a^{-1}$ we extract the masses presented in  table
\ref{table_calc}. These masses are computed in the $\MSB$ scheme (3 GeV); the
conversion from RI-MOM to $\MSB$ is obtained  by using formulae involving the
four-loop anomalous dimensions of the quark mass~\cite{Chetyrkin}.

\subsection{Pseudocritical $\kappa$: $\kappa_{pc}$}
\label{pc}
For any parameter set ($\beta, \kappa_{\rm sea}$), having computed 
the valence masses for several values of $\kappa_{\rm val}$
we extrapolate to a vanishing valence mass. We call 
``pseudocritical $\kappa$'',
$\kappa_{pc}(\kappa_{\rm sea})$, the value of
$\kappa_{\rm val}$ for  which $m_{\rm val}=0$.
The values of $a m_{\rm val}$ as a function of $1/\kappa_{\rm val}$ are 
perfectly compatible with  linear fits.

The pseudocritical $\kappa$'s as a function of $1/\kappa_{\rm sea}$
are also compatible with a linear fit except for one point at 
$\beta=6.0$: $\kappa_{\rm sea}= 0.1510$. 
We did not succeed to understand the reason for this unusual behaviour and have
for the moment withdrawn this point from our fit for $\beta=6.0$.
We call ``critical $\kappa$'' ($\kappa_c(\beta)$) for one $\beta$ the value of 
$\kappa_{\rm sea}$
at which the extrapolated  $\kappa_{pc}$ is equal to $\kappa_{\rm sea}$:
$\kappa_{pc}(\kappa_c)=\kappa_c$.

Our results for $\kappa_c$ are the following:
\beq
\kappa_c(5.6)= 0.158480 (32),\qquad 
\kappa_c(5.8)= 0.154682 (34),\qquad 
\kappa_c(6.0)= 0.152012 (34)
\label{kappac}
\eeq
At $\beta=5.6$, we made the extrapolation using two runs performed by us,
($\kappa_{\rm sea}=0.1575$ and $\kappa_{\rm sea}=0.1560$ ($24^4$) ) and a third
one at $\kappa_{\rm sea}=0.1580$ taken from \cite{FAC}. The value of
$\kappa_c(5.6)$ we obtained is perfectly compatible with the one
($\kappa_c(5.6)=0.15846(5)$) published by SESAM \cite{SESAMold}. Replacing our
run on $24^4$ by the one on $16^4$ induces no significant difference. To our knowledge, the last two values are new.
 
\subsection{Some tests of finite volume effects}
It is clear that a critical point in everything we report here is the risk that
finite volume effect might spoil our results. In our mind the present work should
be mainly a preparation for similar runs on larger volumes ($24^4$) and we want
to be sure that the information gathered on $16^4$ is  relevant enough to tune
our parameters for a  larger volume. We performed two checks with this purpose.

The first one is the comparison 
of the two runs at $\beta = 5.6, \kappa_{\rm sea} = 0.1560$
reported in table~\ref{table_calc}. It can be seen that there is no
significant difference between the results for $16^4$ and $24^4$.

The second one relies on the idea that there could be some kind of  first order
phase transition at very small volume, a deconfinement and/or chiral restoration
transition. Chiral symmetry restoration has the effect of eliminating the
Goldstone boson and thus of invalidating the relation $m_P^2 \propto (m_q +
m_{\bar q}) $ where $m_P$ is the lightest pseudoscalar meson ``mass''$^7$ and
$m_q$ ($m_{\bar q}$) the (anti)quark mass.

\begin{figure}[hbt]
\begin{center}
\mbox{\epsfig{file=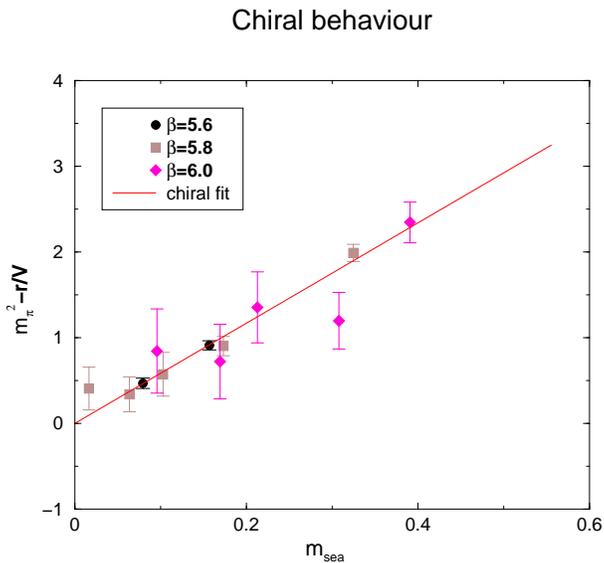,height=7.5cm}} 
\end{center}
\caption{Finite size effects on chiral behaviour.}
\label{gold}
\end{figure}

  Our analysis has found empirically that all our lattice  data\footnote{We included the run
  at $\beta=6.0,  \kappa_{\rm sea}=0.1510$ in this analysis to test in a
  different way its chiral behaviour; however it did not exhibit any pathology
  here.} can  be fitted to a good accuracy by the following formula:
 \beq
 m_P^2 = 2 B m_{\rm sea} + \frac {r}{V} 
 \label{chi}\eeq
Here $m_q=m_{\bar q}$ is the dynamical-quark mass, and $V$ the lattice volume, 
both expressed in physical units. We still take the quark mass renormalised in
the $\MSB$ scheme at 3 GeV. We then obtained   
 \beq
 B = 2.74 (5) \,{\rm GeV}, \qquad r= 1.41 (5){\rm GeV}^{2}\, {\rm fm}^4;
 \label{chiral}\eeq
from a best fit with a $\chi^2/{\rm d.o.f.}=0.57$ (see F{\scshape
ig}.~\ref{gold}). 
 
Eq. (\ref{chiral}) shows a strong but smooth finite volume effect\footnote{We
cannot compare our finite volume correction in eq. (\ref{chi}) to  existing
theoretical studies of finite volume effects on ground state energies \cite{LL}:
indeed, as already stressed, $m_P$ is not really a ground state energy.}, without
any sign of a sudden change of regime, as would be the case with a first order
phase transition.
 
In the infinite volume limit we should recover the pseudoscalar mass
$m_{P,\infty}$. Indeed we checked for $\beta=5.6$ and $\kappa_{\rm sea}=0.1575$
that $m_{P,\infty}^2 \simeq 0.432(18)~{\rm GeV}^2$, in good agreement with SESAM
\cite{SESAMold}: $m_{P}^2 \simeq 0.432(9)~{\rm GeV}^2$. Furthermore
$m_{P,\infty}^2=2.74 (m_q+m_{\bar q})$, which from the pion mass gives
$(m_u+m_d)/2 \simeq 3.6$ MeV and from the kaon mass $m_s
\simeq 90$ MeV at 3 GeV. This compares fairly well to other lattice estimates.

\section{Second iteration: Fitting $\Lambda_{\rm QCD}$ and power corrections}
\label{second}

\subsection{Fitting $\Lambda_{\rm QCD}$ and $O(1/\mu^2)$ coefficient} 

Once we have an approximate estimate of the lattice spacings and dynamical
masses, we now proceed with a combined fit of $\alpha_s$  on the  line with
approximatively constant dynamical-quark mass which goes through $\beta=5.6,
\kappa_{\rm sea} = 0.1560$: $\beta=5.8, \kappa_{\rm sea} = 0.1525$ and
$\beta=6.0, \kappa_{\rm sea} = 0.1505$. This allows to reach momenta as large as
$\sim$ 7.0 GeV, large enough to see the asymptotic behaviour, provided we take
into account $O(1/\mu^2)$ corrections. 

According to our ansatz (\ref{LaFor}), we need  to fit simultaneously the lattice spacing ratios, 
and the parameters  $\Lambda_{\rm
QCD}$ and $c$ (coefficient of $\alpha_{\rm s,pert}/\mu^2$). To fit the lattice 
spacings one needs some analytic function to interpolate between the measured
points and to adjust its parameters simultaneously with  the lattice spacings to
the smallest $\chi^2$. Two approaches are possible. 

One approach is to use the asymptotic four loops
behaviour plus $\alpha_{\rm s,pert}/\mu^2$ corrections  as the analytic function.
This would allow to reach both goals with one stroke but at the expense of eliminating
about one half of the points at $\beta=5.6, \kappa_{\rm sea} = 0.1560$ which are too 
low in energy to follow the asymptotic behaviour.

\begin{figure}[hbt]
\begin{center}
\begin{tabular}{cc}
\hspace{-0.65cm} \mbox{\epsfig{file=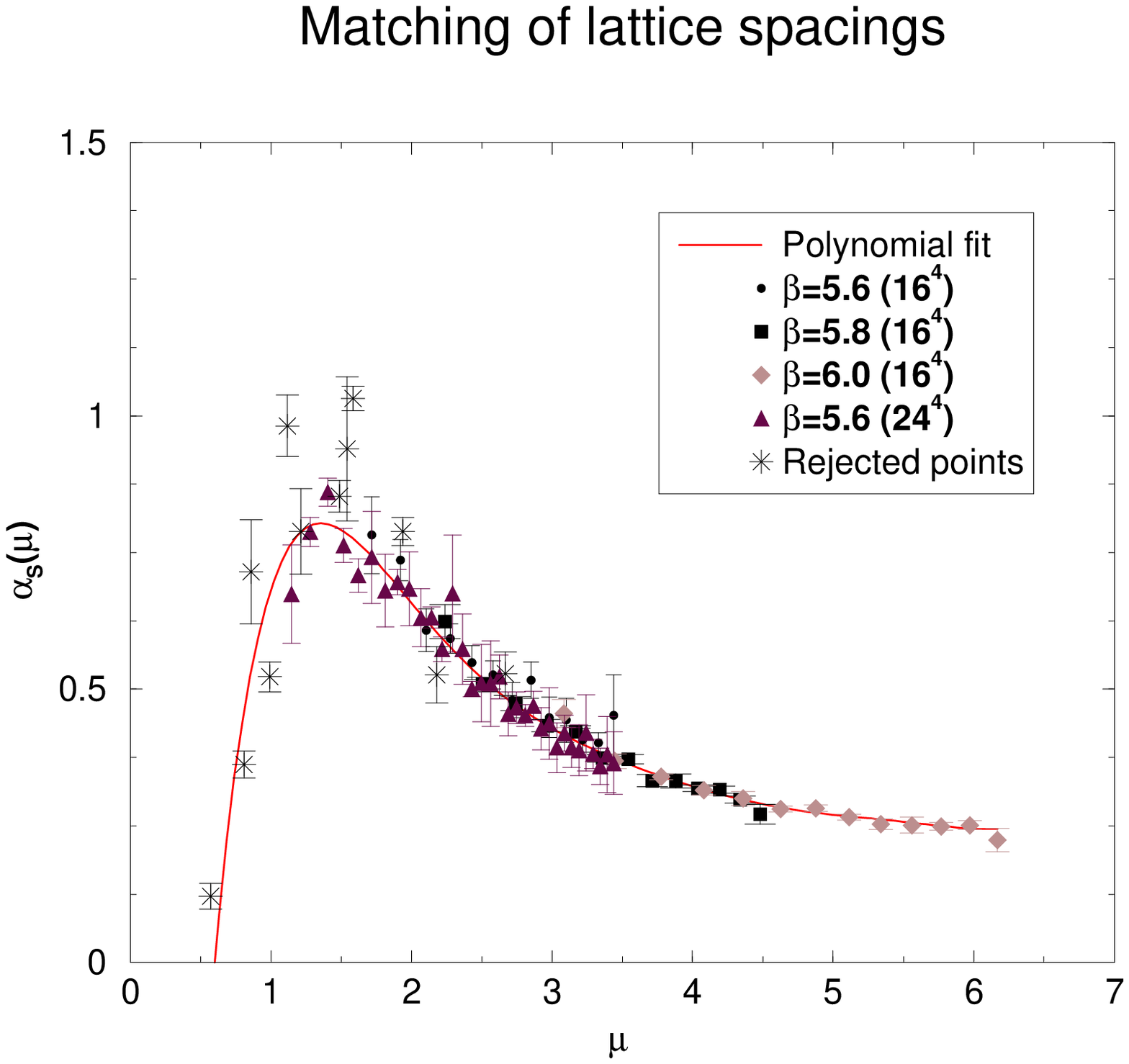,height=7.5cm}} \hspace{0.35cm} & 
\hspace{0.5cm}
\mbox{\epsfig{file=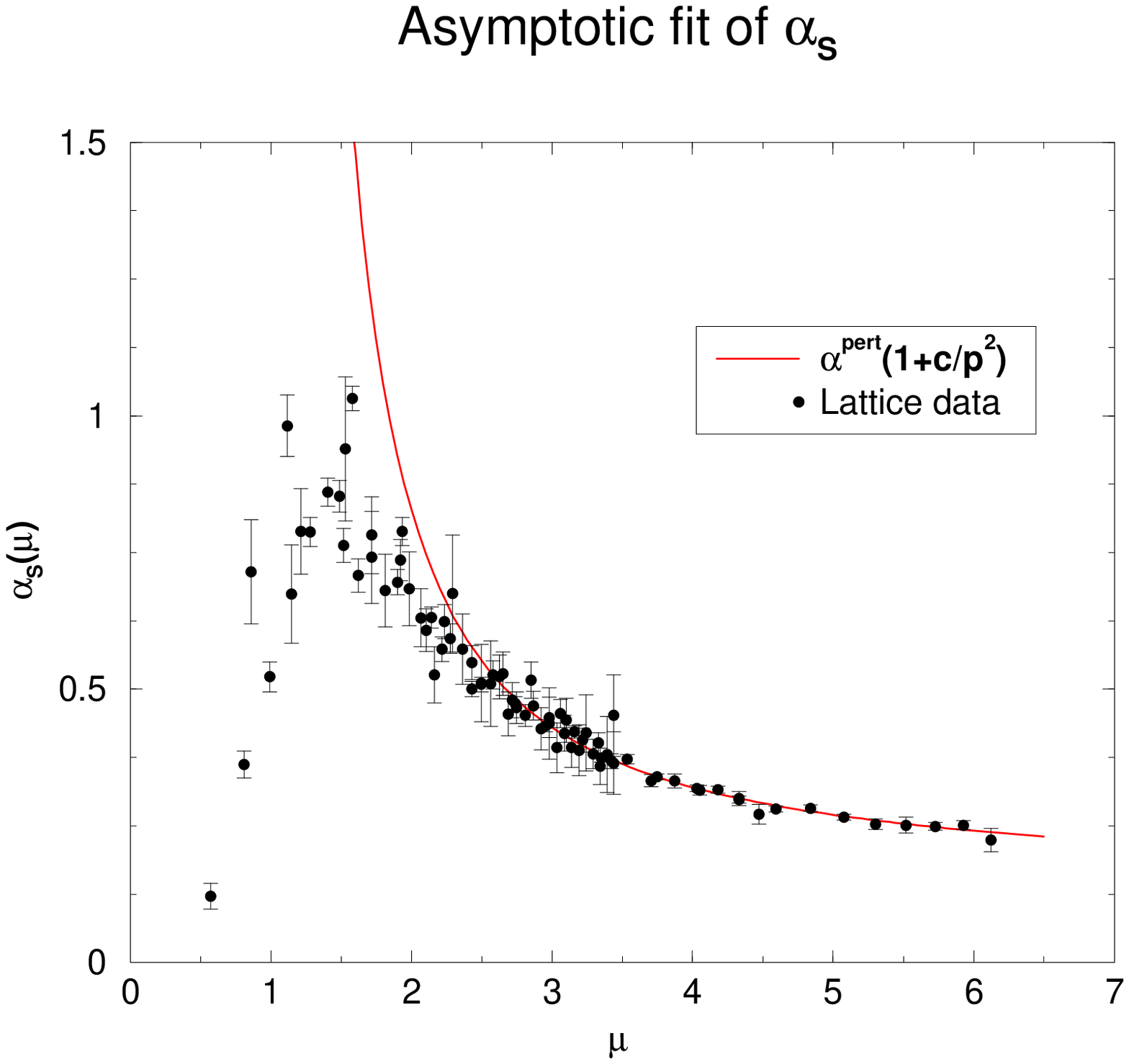,height=7.5cm}} 
\end{tabular}
\end{center}
\caption{Fits of $\alpha_s$ obtained with four lattices. }
\label{Comp}
\end{figure}

The other approach proceeds in two steps as follows. To fit the lattice spacings we have used
polynomials. At $\beta=5.6, \kappa_{\rm sea} = 0.1560$ we have used both the
$16^4$ and the $24^4$ lattices. A universal polynomial (F{\scshape
ig}.~\ref{Comp}, \textit{Matching of lattice spacings}) fitting all points of
the  four lattice settings considered does indeed exist except for a few points
which happen to correspond to $n=(L\mu)^2/(4 \pi^2) \lwrsim 2-4$ where $L$ is the
length of the lattice. We attribute this behaviour to  a strong
finite volume effect \cite{nous} and exclude the points below some IR
cutoff. Varying this IR cut  from $n>2$ to $n>4$ leads to a variation in $\chi^2$ from
$\chi^2/d.o.f. = 1.06$  to $\chi^2/d.o.f. = 0.79$. For lower IR cut-offs the $\chi^2$ increases
dramatically, while for higher IR cut-offs too many points are excluded. The uncertainty induced by the choice of the cutoff
is taken into account in the systematic error which affects the values quoted below.   At this point, it should perhaps be emphasized again that this procedure was tested on quenched data, providing us with the generally admitted lattice spacing ratios. In this way we obtain:
\beq 
\begin{array}{c}
a^{-1}(5.8, 0.1525) = 2.85 \pm .09 \pm .04 \times  \frac
{a^{-1}(5.6,0.1560)}{2.19\,{\rm GeV}} \ {\rm GeV} \\
a^{-1}(6.0, 0.1505) = 3.92 \pm .11 \pm .07 \times  \frac
{a^{-1}(5.6,0.1560)}{2.19\,{\rm GeV}} \ {\rm GeV}
\end{array}
\label{mailles}
\eeq
where the central value  corresponds to a cut at $n \ge 3$,  the first error is statistical and the second is  systematic. 
From now on, we are going to use these
ratios,  correcting the first iteration estimates shown in table
\ref{table_calc}.  

\vskip 0.5 cm
\begin{table}
\begin{center}
\begin{tabular}{|c|c|c|c|}\hline
$\mu_{\rm min}$\,(GeV) & $\Lambda_{\rm \overline{MS}}$\, (MeV)& $c$ \,(GeV$^2$) & $\chi^2$/d.o.f.
\\ \hline
2.6 & 264(24) & 2.66(77)& 0.58\\
3.1 & 256(20) & 3.03(85) & 0.56\\
3.6 & 267(29) & 2.51(1.21) & 0.54\\
4.1 & 269(29) & 2.44(1.75) & 0.71\\
 \hline
\end{tabular}
\vskip 1.5 cm
\caption{Four-loop fit with power corrections, eq.  (\ref{LaFor}), on
varying energy windows ($>\mu_{\rm min}$).  The stability of the fit is
fairly good.  
  }
\label{table_lambda}
\end{center}
\end{table}

Once the lattice spacings have been estimated we 
perform a combined fit of
 $\Lambda_{\rm \overline {MS}}^{N_f=2}$ and the coefficient 
 $c$ as defined in eq.  (\ref{LaFor}) with
 $\alpha_{\rm s,pert}^{\rm N_f=2}$ given by the r.h.s.
 of eq. (\ref{3lp}). The result is plotted in
 F{\scshape ig}.~\ref{Comp}, (\textit{Asymptotic fit of $\alpha_s$}).

From the results in table \ref{table_lambda} we conclude:

\beq \Lambda_{\rm \overline{MS}}^{N_f=2} = 264(27)\frac
{a^{-1}(5.6,0.1560)}{2.19\,
{\rm GeV}} {\rm MeV } \qquad c = 2.7(1.2) \left[\frac {a^{-1}(5.6,0.1560)}{2.19\
{\rm GeV}} {\rm GeV}\right]^2,\label{lambdanf0} \eeq

The same analysis using the formula in 
eq.  (\ref{LaFor}) 
leads to 
\beq
\Lambda_{\rm \overline{MS}}^{N_f=0} \simeq 252(10) {\rm MeV }
\qquad c =1.0(1) {\rm GeV^2}\label{lambdanf2}
\eeq
from our $N_f=0$ data \footnote{If the fit for $N_f=0$ is performed according to
$\alpha_s^{\rm Latt}(\mu^2)=\alpha_{\rm s,pert} (\mu^2)\,+ \frac{c}{\mu^2}$  
instead of eq. (\ref{LaFor})  the result is $\Lambda_{\rm \overline{MS}}^{N_f=0}
\simeq 237(10) {\rm MeV } $ \cite{direnzo}.} \cite{nous}. 

\begin{figure}[hbt]
\begin{center}
\epsfig{file=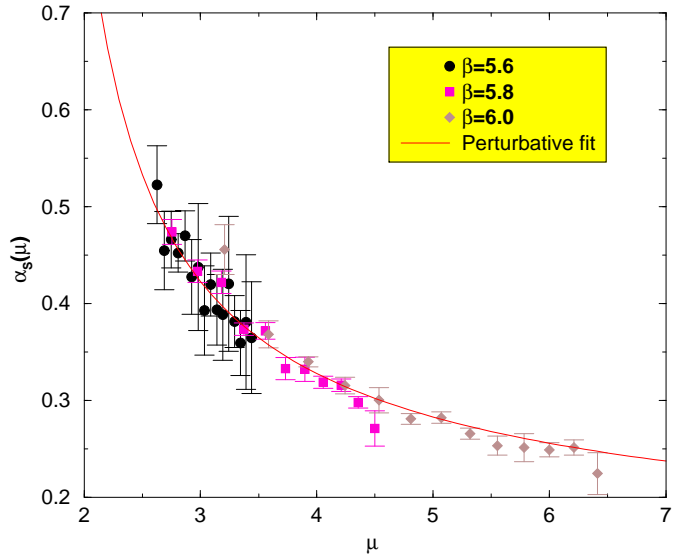,height=7.5cm} 
\caption{Purely perturbative fit of $\alpha_s$. }
\label{claudefig}
\end{center}
\end{figure}

If we use the first approach mentioned above, i.e. fitting from the beginning with the formula of
eq. (\ref{LaFor}), the result are perfectly compatible with eqs. (\ref{mailles}) and (\ref{lambdanf0}).
If we try the same procedure without power corrections,
keeping an energy window ranging from 2.6~GeV, we can obtain a best fit with the following parameters (see F{\scshape ig}.~\ref{claudefig}):
\beq 
\begin{array}{c}
a^{-1}(5.8, 0.1525) = 2.86(6) \times \frac{a^{-1}(5.6,0.1560)}{2.19\,{\rm GeV}} \ {\rm GeV} \\
a^{-1}(6.0, 0.1505) = 4.08(9) \times \frac{a^{-1}(5.6,0.1560)}{2.19\,{\rm GeV}} \ {\rm GeV}
\end{array}
\label{claude}
\eeq
and
\beq
\Lambda_{\rm \overline{MS}}^{N_f=2} = 345(6) \qquad \chi^2/d.o.f. = 0.96
\label{lambda_claude}
\eeq
It is not surprising that $\Lambda_{\rm \overline{MS}}^{N_f=2}$ 
is larger when the fit does not include $1/\mu^2$ corrections, since
the parameters $\Lambda_{\rm \overline{MS}}^{N_f=2}$ and $c$ vary naturally  a contrario: if one increases the other
decreases. There is a possible contradiction between the acceptable $\chi^2$
found in (\ref{lambda_claude}) and the fact that in eq. (\ref{lambdanf0}) the coefficient
$c$ is three standard deviations away from 0. This might be due to some
correlations in the data; further study is needed to settle the origin of the discrepancy. 

  Let us 
nevertheless underline three facts that make us confident about the pertinence of
incorporing power corrections: \begin{enumerate} \item The OPE analysis
\cite{nouspdeux} still holds in the unquenched case, so we theoretically expect
the presence of power corrections; we wonder about their magnitude. \item In the
pure Yang-Mills case \cite{nous,direnzo,nouspdeux} where high accuracy
computations were achieved, the purely perturbative formula could not provide us
with a good fit, but the one including power corrections managed to. This
``numerical evidence'' is comforted by the good agreement of our value of
$\Lambda_{\rm \overline{MS}}^{N_f=0}$ and that of the ALPHA collaboration
\cite{nouspdeux,sommer}. \item Provided we do control the systematic
uncertainties, the current data already show a tendancy to discriminate in favour
of the inclusion of non-perturbative terms. Indeed, in the case of a fit\emph{without} power corrections, the lattice data are lying
\emph{over} the fitting curve for the lowest values of our energy window, and
\emph{under} the curve for the highest energy values. We expect this tendancy to become clearer with higher
statistics. \end{enumerate}

Of course, we are conscious that the present letter would benefit from a deeper
discussion of systematic errors. In particular the use of non-improved dynamical
quarks leads to $O(a)$ errors.  However our lattice spacings are all rather small and these errors partly cancel in the ratios of lattice spacings.  The dominant error
is thus an overall  $O(a)$ error on the calibrating lattice spacing.  This error
propagates multiplicatively to the values  of $\Lambda_{\rm
\overline{MS}}^{N_f=2}$ and c. We are not in a position at the moment to estimate in a
reliable way this systematic error. A rough estimate can be obtained either by
analogy with the quenched case, looking at Fig.~4 in  \cite{edwards} for
$a^{-1}\sim 2.2$ GeV or directly in the unquenched case from Fig.~2 in
\cite{SESAMold}.  A crude estimate is 20\%. This would give 50 MeV on 
$\Lambda_{\rm \overline{MS}}^{N_f=2}$. The effect on $\alpha_s(M_Z)$ will be a
systematic error of $\pm .004$.

It is useful to notice that the $O(a)$ errors are $O(am_{\rm sea})$ or
$O(a \Lambda_{\rm QCD})$. No errors $O(a\mu)$ are expected due to symmetry
reasons: the hypercubic symmetry of the lattice ($\mu_\nu \rightarrow -\mu_\nu$ for any $\nu$)
implies that the momentum dependent errors are $O(a^2 \mu^2)$, exactly as in the
pure Yang-Mills case. We did not see any significant effect of the latter errors
for the momenta that we have considered.  In particular such errors should show 
up in fig. \ref{Comp} as a systematic deviation from the global fit
for the data with largest $a^2 \mu^2$. 
 
\subsection{Estimating $\alpha_s(M_z)$}

At an energy of the order of the $Z$ meson mass the
$O(1/\mu^2)$ power correction becomes irrelevant.
We will therefore only keep $\alpha_{\rm s,pert}$, the perturbative part 
of $\alpha_s$ from our fit and extrapolate. We 
proceed as indicated in \cite{pdg}. We start from an energy of 1.3 GeV,
the $\overline {\rm MS}$ charm mass which is
taken as the charm threshold\footnote{We have preferred to follow
the tradition here, although it is not clear to us why one should
use the $\overline {\rm MS}$ mass and not the pole mass, and why
the threshold is at $m_c$ and not $2 m_c$ where the charm loop dispersive
contribution starts for the gluon propagator.}. At such an energy
we will extrapolate from our quenched and two-flavour
results to three flavours. We then start evolving
up with four flavours to the beauty threshold,
4.3 GeV,  and then further up with 5 flavours to $M_Z$.

Applying eqs. (\ref{lambda3}-\ref{alpha4}) 
with the values of $\Lambda_{\rm \overline{\rm MS}}$
in eqs. (\ref{lambdanf0}) and (\ref{lambdanf2})
and assuming ${a^{-1}(5.6,0.1560)}={2.19\,{\rm GeV}}$
we get in $\rm \overline {MS}$ scheme
\beq
\alpha_{\rm s,pert}^{\rm N_f=0}(1.3) = 0.259(6),\qquad \alpha_{\rm s,pert}^{\rm N_f=2}(1.3)
= 0.306(20),
\qquad \alpha_{\rm s,pert}^{\rm N_f=3}(1.3) =  0.329(26)
\label{alphafin}\eeq
where the N$_f = 0,2$ results come from direct lattice estimates
in \cite{nous} and in this work, while the   N$_f = 3$
has been extrapolated from the two latter\footnote{We simply assume that
the extrapolation to an odd number of flavors is legitimate, not knowing 
what to do better.}.

The evolution up to $M_Z$ (where non-perturbative corrections are negligeable) 
 and down to $M_{\tau}$ gives
\beq
\alpha_s(M_Z) = 0.113(3)(4) \qquad  \alpha_{\rm s,pert}(M_{\tau}) = 0.283(18)(37)
\label{alphaMZ}
\eeq 
where the second error comes from the systematic error  on the calibrating 
lattice spacing.

\section{Discussion and conclusion}

We should reemphasize that this is mainly a progress report.  Most of the
results reported here were performed on small volumes and with rather large
quark masses.  Our goal was to undertake a first exploration of the
parameter space. It turned out that the results seem to make sense. The
rather smooth junction of the $\alpha_s$ points from three different
lattices  show that overwhelming ultraviolet or infrared lattice artifacts
are absent.

The points from different  lattices with identical momenta do 
coincide unless $(L\mu)^2/(4\pi^2)\lwrsim 2$. Suffering presumably  from
strong finite volume effects these points  have been excluded from the
global fits. The comparison at
$\beta =5.6, \kappa_{\rm sea}=0.1560$ of the $16^4$ and the $24^4$ 
volumes are encouraging and should be extended to other
sets ($\beta, \kappa_{\rm sea}$).
The finite volume effect on masses seems to be well accounted for
by  eqs. (\ref{chi}) , (\ref{chiral}) , and the good agreement of 
$m_{P,\infty}$ with the estimate in \cite{SESAM}, performed on a larger
time interval, confirms this optimism.

Our result for $\alpha_s(M_Z)$ is about 2 standard deviations
 below  the world average experimental
$\alpha_s(M_Z) = 0.119(2)$ \cite{pdg}. It is slightly larger, although
compatible  within errors, with the result\footnote{This results from the fact 
that our value $\Lambda_{\rm \overline{MS}}^{N_f=2}= 264(27)$ is larger than
the value 217(16)(11) from \cite{schierholz}.}   
of \cite{schierholz}: $\alpha_s(M_Z) = 0.1076(20)(18)$. Older results 
using NRQCD were closer to experiment:
$\alpha^{(5)}_{s}(M_Z) = 0.1174(24)$ \cite{davies}, 
$\alpha_{s}^{(5)}(M_Z) = 0.118(17)$ \cite{SESAM}. 
Our result for $\alpha_{\rm s,pert}(M_{\tau})$ is also 2 $\sigma$'s below the
experimental value of 0.334(22) MeV \cite{ALEPH}. However, the meaning of this comparison is unclear because we cannot take into account the non-perturbative contribution to $\alpha_s^{\MSB}$ at $M_{\tau}$.

We consider the fact that our preliminary result is 2 $\sigma$'s below
experiment  as  very encouraging. We should stress  that the error presented in
eq. (\ref{alphaMZ}) corresponds to the statistical  error and only to some
systematic errors: mainly the choice of the fitting window and the calibration
error.   Other systematic effects  should be systematically explored such as 
that of the dynamical-quark action and that of the mass of the  dynamical-quark
(ours are rather heavy). A calculation with a lighter dynamical-quark mass 
is in progress. As a final remark we would like to stress that our
value for $\alpha_s(M_Z)$ is strongly correlated to the  rather large 
$1/\mu^2$ corrections that we find in our fit. Starting from eq. (\ref{claude})
i.e. from a fit without power corrections we obtain $\alpha_s(M_Z) = 0.1211(3)(40) $.
As already stated,  in the fits, $\Lambda_{\rm
\overline{MS}}$ and $c$ show an understandable tendency to vary a contrario.  We are clearly encouraged to follow on this
analysis and try to refine our result for $\alpha_s(M_Z)$.

\section*{Acknowledgements.}

These calculations were performed partly on the QUADRICS QH1 located in the
Centre de Ressources Informatiques (Paris-sud, Orsay) and purchased thanks to a
funding from the Minist\`ere de l'Education Nationale and the CNRS.  We are
indebted to the Lattice group of the University of Rome I for  allowing us to run
a part of it on one of their APE100. We thank Damir Becirevic and Alain Le
Yaouanc for several inspiring comments. This work was supported in part by the
European Union Human Potential Program under contract HPRN-CT-2000-00145,
Hadrons/Lattice QCD. Ph.~B. acknowledges the Physics Department of Rome~I and J.~R.-Q. the LPT of Universit\'e Paris-Sud for hospitality.

\end{document}